\documentclass[lettersize,journal]{IEEEtran}
\usepackage{amsmath,amsfonts}
\usepackage{algorithmic}
\usepackage{algorithm}
\usepackage{textcomp}
\usepackage{stfloats}
\usepackage{url}
\usepackage{verbatim}
\usepackage{graphicx}
\usepackage{cite}
\usepackage{subfigure}
\usepackage[cmintegrals]{newtxmath}
\begin{document}

\title{Compressive Spectrum Sensing with 1-bit ADCs}

\author{Jian Yang,
	Zihang Song,
	Han Zhang,
	and Yue Gao,~\IEEEmembership{Fellow ,~IEEE}
\thanks{This work was supported by the National Key Research and Development Program of China (Grant NO.2022YFB20210044).}
\thanks{Jian Yang (e-mail: yangjian1@caict.ac.cn) is with the Mobile Communications Innovation Center, China Academy of Information and Communications Technology, Beijing 100191, China. Zihang Song is with the Department of Engineering, King's College London, Strand, London, WC2R 2LS, United Kingdom. Han Zhang is with the School of Information and Communication Engineering, Beijing University of Posts and Telecommunications, Beijing 100876, China. Yue Gao is with the Institue of Space Internet and School of Computer Science, Fudan University, Shanghai, China.}}

\markboth{Journal of \LaTeX\ Class Files,~Vol.~14, No.~8, August~2021}%
{Shell \MakeLowercase{\textit{et al.}}: Compressive Spectrum Sensing with 1-bit ADCs}


\maketitle

\begin{abstract}
	Efficient wideband spectrum sensing (WSS) is essential for managing spectrum scarcity in wireless communications. However, existing compressed sensing (CS)-based WSS methods require high sampling rates and power consumption, particularly with high-precision analog-to-digital converters (ADCs). Although 1-bit CS with low-precision ADCs can mitigate these demands, most approaches still depend on multi-user cooperation and prior sparsity information, which are often unavailable in WSS scenarios. This paper introduces a non-cooperative WSS method using multicoset sampling with 1-bit ADCs to achieve sub-Nyquist sampling without requiring sparsity knowledge. We analyze the impact of 1-bit quantization on multiband signals, then apply eigenvalue decomposition to isolate the signal subspace from noise, enabling spectrum support estimation without signal reconstruction. This approach provides a power-efficient solution for WSS that eliminates the need for cooperation and prior information.
\end{abstract}


\section{Introduction}
{Wideband} spectrum sensing (WSS) is crucial for enabling secondary users (SUs) to identify available spectrum in cognitive radio (CR) systems \cite{10261286,8882268,song2021survey,10461095}. To capture wideband signals accurately, traditional WSS methods rely on high sampling rates, which significantly increase power consumption and data processing requirements. Compressed sensing (CS) has been introduced to WSS to reduce the sampling rate by leveraging the sparsity of primary users (PUs) across the frequency range, enabling sub-Nyquist sampling \cite{song2024gbsense,9454267,9971798,song2022approaching}. However, most CS-based WSS methods use high-precision analog-to-digital converters (ADCs), which, though effective, impose considerable power and data handling costs on the SUs.

To overcome this, 1-bit CS has recently been proposed, using 1-bit ADCs that capture only the sign of each measurement, dramatically reducing both power consumption and data rate \cite{10497157}. For instance, a 12-bit ADC operating at 3.2 GSps consumes approximately 3.8 Watts, while a 1-bit ADC at the same rate uses only 20 $\mu$Watts \cite{TI,8864102}. This makes 1-bit CS an attractive solution for WSS. However, applying 1-bit CS to WSS introduces new challenges: the loss of amplitude information complicates signal reconstruction, and existing methods often require prior knowledge of the signal’s sparsity order, which is typically unknown in WSS scenarios and difficult to estimate due to significant quantization noise from 1-bit ADCs \cite{9597562,8682936}. Furthermore, many existing 1-bit CS-based WSS approaches rely on multi-user cooperation to enhance sensing performance and achieve reliable spectrum occupancy detection \cite{7906062,Shengnan,6418031,6178284}.

To address these limitations, this paper proposes a novel non-cooperative WSS method that leverages 1-bit ADCs within a multicoset sampling framework. Our approach eliminates the need for high-precision ADCs, multi-user cooperation, and prior sparsity information. The primary contributions of this paper are as follows:

\begin{enumerate}
    \item We analyze the effects of 1-bit quantization on multiband signal models in the frequency domain, adapting this analysis specifically to multicoset sampling.
    \item By employing eigenvalue decomposition, we extract the signal subspace from environmental and quantization noise, allowing us to accurately estimate the support of the occupied spectrum without requiring full signal reconstruction.
    \item This approach enables standard CS reconstruction algorithms to function effectively without prior sparsity knowledge, making it a low-power and efficient solution for WSS in CR systems.
\end{enumerate}

Our approach provides a low-power alternative to conventional CS-based WSS, eliminating the need for high sampling rates, multi-user cooperation, and prior information on signal sparsity.

The rest of this paper is organized as follows. In Section II, the WSS frame and multicoset sampling system are presented. In Section III, the effect of quantization on the multiband signal model is analyzed and the subspace-aided compressive spectrum sensing method is proposed. Simulation results are given in Section IV.

\section{System Model}
\label{s2}

In this section, we introduce the signal model for WSS and the multicoset sampling system used in this process.

\subsection{Signal Model}
In WSS, the SUs monitor the radio spectrum within the frequency range $[0, W]$, where there are $K$ PU transmissions. The CR system divides this spectrum into $L$ channels, each with a bandwidth of $B = W/L$. Without loss of generality, we assume that each PU occupies a single channel. The channels are indexed from $1$ to $L$, and the active channels occupied by PUs form a support set, denoted by $\mathcal{S} = \{s_1, s_2, \ldots, s_K\}$. The sparsity order of the signals, defined as $K = |\mathcal{S}|$, corresponds to the number of active PUs. The primary goal of WSS is to estimate this support set $\mathcal{S}$.

During the spectrum sensing phase, all CR devices are silent, as enforced by media access control layer protocols. Therefore, the received signal by CR devices consists only of the signals transmitted by active PUs and additive white noise. The complex baseband model for the received signal at the SU can be represented as:
\begin{equation}\label{e111}
	x(t) = \sum_{i=1}^{K} s_i(t) + n(t),
\end{equation}
where $s_i(t)$ represents the signal from the $i$-th PU, and $n(t) \sim \mathcal{CN}(0, \sigma_n^2)$ is circularly symmetric complex Gaussian noise with zero mean and variance $\sigma_n^2$. The signal $x(t)$ spans the frequency band $[0, W]$.

The Fourier Transform (FT) of the received signal $x(t)$ is given by:
\begin{equation}\label{e112}
	X(f) = \int_{-\infty}^{+\infty} x(t) e^{-j 2 \pi f t} \, dt = S(f) + N(f),
\end{equation}
where $S(f) = \sum_{i=1}^{K} S_i(f)$ denotes the FT of the PUs' signals, and $N(f)$ represents the FT of the noise. Throughout this analysis, we assume that SUs have no prior information about the PUs' signals, and that each PU's signal $s_i(t)$ is independent of the others.

\subsection{Multicoset Sampling System}

\begin{figure}[!t]
	\centering
	\includegraphics[width=0.8\columnwidth]{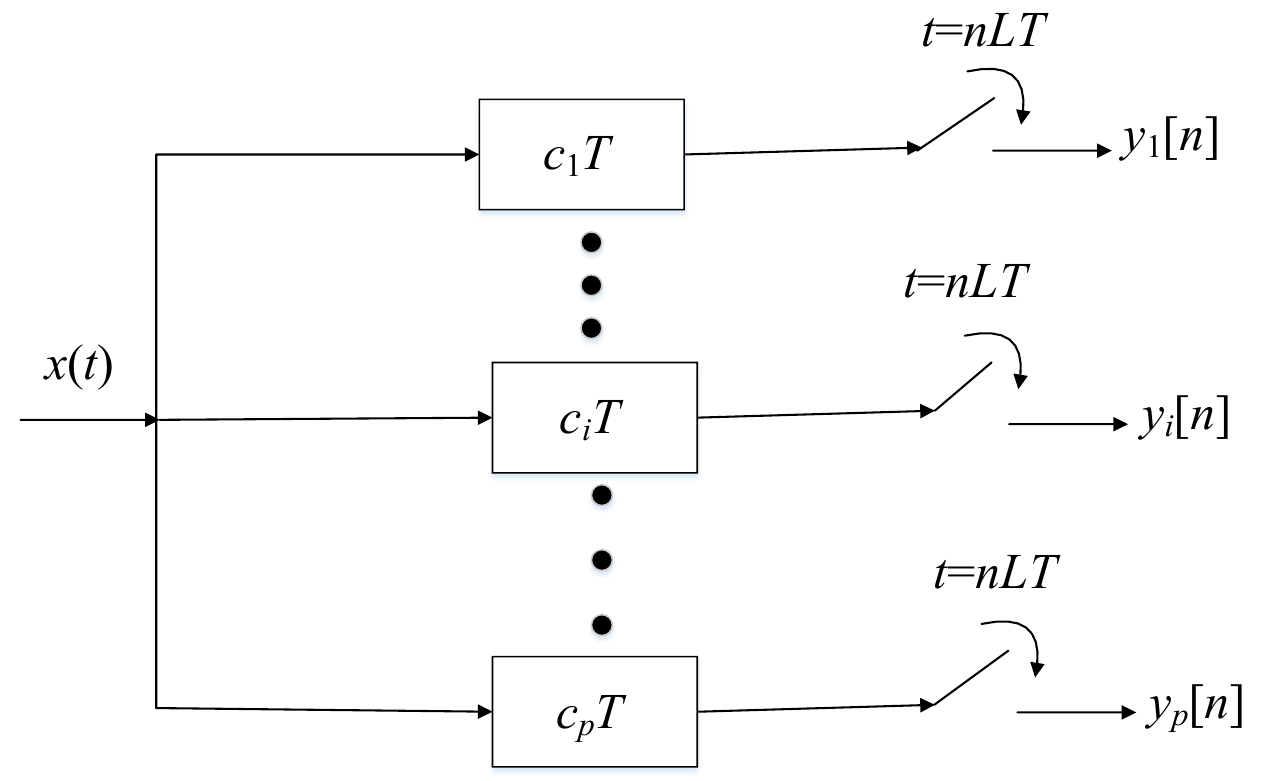}
	\caption{The parallel implementation of multicoset system.}
	\label{fig:multicoset}
\end{figure}

With a fixed number of channels $L$ and a spectrum bandwidth of $W$, the multicoset sampler acquires non-uniform samples at time instants $t=(mL+c_i)T$, where $i=1, 2, \ldots, p$, $m \in \mathbb{Z}$, $c_i$ is an integer, and $T=1/W$ is the Nyquist sampling interval. In practice, the multicoset sampler can be implemented using $p$ parallel cosets, with each coset sampling the signal uniformly at a sampling rate of $f_s=W/L$ and with a time delay of $c_iT$. The set $\mathcal{C} = \{c_i\}_{i=1}^p$ consists of $p$ distinct integers selected from $\{0,1,2,\ldots,L-1\}$ as in \cite{song2024nonuniform}, as shown in Fig. \ref{fig:multicoset}. For the $i$-th coset, the sample sequence can be expressed as 
\begin{equation}\label{e222}
	y_i[m] = x\left((mL+c_i)T \right), \quad m\in \mathbb{Z}.
\end{equation}
The discrete-time Fourier transform (DTFT) of $y_i[m]$ is
\begin{equation}\label{e1}
	\begin{aligned}
		Y_i\left(e^{j2\pi fLT}\right) &= \sum_{m=-\infty}^{+\infty} y_i[m]e^{-j2\pi fmLT}\\
		&=\frac{1}{LT}\sum_{n=-\infty}^{+\infty} X\left(f-\frac{n}{LT}\right)e^{j2\pi \left(f-\frac{n}{LT}\right)c_iT}\\
		&=\frac{1}{LT}e^{j2\pi fc_iT}\sum_{n=-\infty}^{+\infty} e^{-j2\pi n \frac{c_i}{L}} X\left(f-\frac{n}{LT}\right).
	\end{aligned}
\end{equation}
Since $x(t)$ is band-limited within the frequency range $[0,W]$, the DTFT of $y_i[n]$ can be expressed as
\begin{equation}\label{e2}
	Y_i\left(e^{j2\pi fLT}\right) =\frac{1}{LT}e^{j2\pi fc_iT}\sum_{n=0}^{L-1} e^{-j2\pi n \frac{c_i}{L}} X\left(f-\frac{n}{LT}\right)
\end{equation}
for $f \in [0, B]$. This shows that $Y_i\left(e^{j2\pi fLT}\right)$ is a weighted sum of all channel spectra. 

Define a vector $\mathbf{Y}(f)$ with $p$ entries $Y_i(f) = LTe^{-j2\pi fc_iT}Y_i \left(e^{j2\pi fLT}\right)$. Then, the DTFT of the sample sequences from all the cosets can be unified in matrix form as
\begin{equation}\label{e22}
	\mathbf{Y}(f) = \mathbf{A}\mathbf{X}(f),
\end{equation}
where $\mathbf{X}(f)$ is an $L \times 1$ vector with entries $X_j(f) = X\left(f-\frac{j-1}{LT}\right)$ for $j=1,2,\ldots,L$, and $\mathbf{A}$ is a $p \times L$ matrix with entries $A_{ij} = e^{-j2\pi (j-1) \frac{c_i}{L}}$ for all $i=1,2,\ldots,p$ and $j=1,2,\ldots,L$.

Since $\mathbf{A}$ is a sub-matrix of the Fourier matrix, based on CS theory, $\mathbf{X}(f)$ can be reconstructed using certain CS algorithms with high probability when $p > \omega K \log(L)$, where $\omega$ is a constant \cite{bandeira2018}.

\section{Subspace-Aided Compressive Spectrum Sensing for Multicoset with 1-bit ADCs}

In this section, we introduce 1-bit ADCs into the multicoset sampling system. First, we analyze the effect of quantization noise on the signal model. Then, we propose a novel multicoset sampling scheme that incorporates 1-bit ADCs. Finally, we present a subspace-aided compressive spectrum sensing algorithm designed to address the impact of both quantization noise and Gaussian noise.

\subsection{Effect of 1-bit ADCs on the Signal Model}

Let $z(t) = \mathcal{Q}(x(t))$ represent the 1-bit quantized version of the signal $x(t)$, with $Z(f)$ as the Fourier Transform (FT) of $z(t)$. The function $\mathcal{Q}(\cdot)$ denotes the 1-bit quantization operation, which preserves only the sign information of the signal, defined as:
\begin{equation}\label{e33}
	\mathcal{Q}(x(t)) \coloneq \frac{1}{\sqrt{2}} \left( \text{sign}(\Re\{x(t)\}) + j \, \text{sign}(\Im\{x(t)\}) \right),
\end{equation}
where $\Re\{x(t)\}$ and $\Im\{x(t)\}$ are the real and imaginary components of the signal $x(t)$, respectively. 

Fig. \ref{1bitSignal} shows an example of $X(f)$ and the spectrum of its 1-bit quantized counterpart, $Z(f)$, for a scenario with 10 channels, where 4 channels are occupied. As illustrated, 1-bit quantization introduces noise that spreads across the entire frequency spectrum, making it challenging to directly estimate the sparsity order of the signal.

\subsection{Multicoset with 1-bit ADCs}
To reduce the power consumption of WSS, we propose a novel multicoset sampling scheme using 1-bit ADCs. In this scheme, the sample sequence of the $i$-th coset with 1-bit ADCs is given by
\begin{equation}\label{e3}
	q_i[m] = \mathcal{Q}(y_i[m]),
\end{equation}
where $\mathcal{Q}(\cdot)$ denotes the 1-bit quantization function. The discrete-time Fourier transform (DTFT) of $q_i[m]$ is then expressed as
\begin{equation}\label{e4}
	Q_i(e^{j2\pi fLT}) = \frac{1}{LT}e^{j2\pi fc_iT}\sum_{n=0}^{L-1} e^{-j2\pi n \frac{c_i}{L}} Z\left(f-\frac{n}{LT} \right),
\end{equation}
for $f \in [0, B]$. Formula (\ref{e4}) shows that the spectrum $Q_i(e^{j2\pi fLT})$ is a weighted sum of the spectra of all channels, each quantized by a 1-bit ADC.

\begin{figure}[!t]
	\centering
	\includegraphics[width=1\columnwidth]{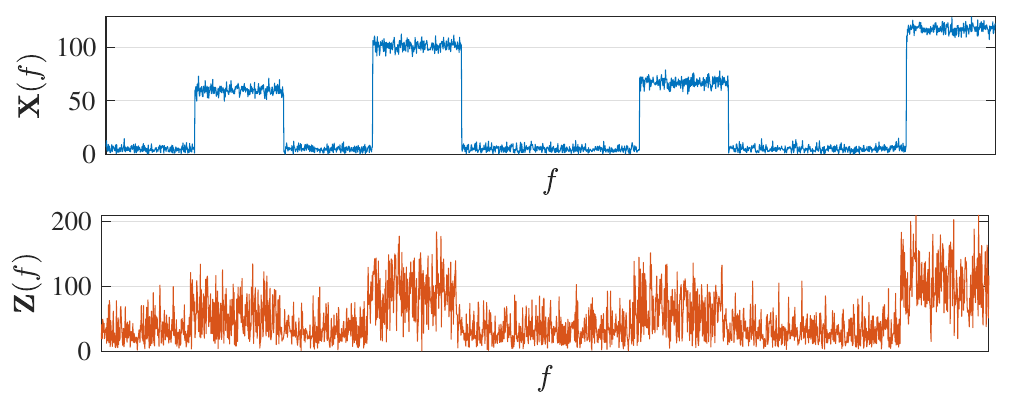}
	\caption{An example of $\mathbf{X}(f)$ and the spectrum of its 1-bit quantized samples $\mathbf{Z}(f)$ with 10 channels in which 4 channels are occupied.}
	\label{1bitSignal}
\end{figure}

Similar to formula (\ref{e22}), we define a vector $\mathbf{Q}(f)$ with entries $Q_i(f) = LTe^{-j2\pi fc_iT}Q_i(e^{j2\pi fLT})$. The DTFT of samples from all cosets can then be unified in matrix form as
\begin{equation}\label{e5}
	\mathbf{Q}(f) = \mathbf{A}\mathbf{Z}(f),
\end{equation}
for $f \in [0, B]$, where $\mathbf{Z}(f)$ is an $L \times 1$ vector with entries $Z_j(f) = Z\left(f - \frac{j-1}{LT}\right)$ for $j = 1, 2, \ldots, L$.

From Fig. \ref{1bitSignal}, we observe that while the vector $\mathbf{X}(f)$ is jointly sparse for $f \in [0, B]$, the quantized vector $\mathbf{Z}(f)$ is not jointly sparse due to the random nature of quantization noise. This randomness in quantization noise disrupts the sparsity of the quantized signal’s spectrum, which complicates the application of traditional CS reconstruction methods.

\subsection{Subspace-Aided Compressive Spectrum Sensing method}
Since 1-bit ADCs lead to serious quantization noise, we propose a novel subspace-aided compressive spectrum sensing algorithm to reduce the effect of the quantization noise on WSS. 

Although $X(f)$ is heavily distorted by 1-bit quantization, fortunately, the Bussgang theorem \cite{demir2020bussgang} states that there is a linear relationship between the covariance of the input and the output of a nonlinear system. Based on the Bussgang theorem, the 1-bit samples of multicoset can be linearized as
\begin{equation}\label{e7}
	q_i[m] = \mathcal{Q}(y_i[m]) = By_i[m]+\psi_i[m],
\end{equation}
where $\psi_i[m]$ is the non-Gaussian distortion caused by 1-bit quantization that is uncorrelated to $y_i[m]$, and 
\begin{equation}\label{e8}
	B \triangleq \frac{\mathbb{E}\{\mathcal{Q}(y_i[m])y_i[m]\}}{\mathbb{E}\{y_i^2[m]\}}
\end{equation}
is the  Bussgang gain. Therefore, the DTFT of $q_i[m]$ can be denoted by
\begin{equation}\label{e9}
	Q_i(e^{j2\pi fLT}) = BY_i(e^{j2\pi fLT}) + \Psi_i(e^{j2\pi fLT}),
\end{equation}
where $\Psi_i(e^{j2\pi fLT})$ is the DTFT of $\psi_i[m]$. 

Define $\mathbf{S}(f) = [S(f),S(f-\frac{1}{LT}),S(f-\frac{2}{LT}),\cdots,S(f-\frac{L-1}{LT})]^T$, $\mathbf{N}(f) = [N(f),N(f-\frac{1}{LT}),N(f-\frac{2}{LT}),\cdots,N(f-\frac{L-1}{LT})]^T$ and $\Psi_i(f) = LTe^{-j2\pi fc_iT}\Psi_i(e^{j2\pi fLT})$. We can rewrite (\ref{e5}) as
\begin{equation}\label{e10}
	\begin{aligned}
		\mathbf{Q}(f) =& B\mathbf{Y}(f)+\mathbf{\Psi}(f)\\ 
		=& B\mathbf{A}\mathbf{X}(f)+\mathbf{\Psi}(f)\\
		=& B\mathbf{A}(\mathbf{S}(f)+\mathbf{N}(f))+\mathbf{\Psi}(f),
	\end{aligned}
\end{equation}
where $\mathbf{\Psi}(f)$ is a vector with entries $ \Psi_i(f),i=1,2,\cdots,p$.

The auto-correlation matrix of $\mathbf{Q}(f)$ can be denoted by
\begin{equation}\label{e11}
	\mathbf{R}_Q = B^2\mathbf{A}\mathbf{R}_S\mathbf{A}^H + \frac{B^2\sigma_n^2}{LT^2}\mathbf{I}+\mathbf{R}_\Psi,
\end{equation}
where $\mathbf{R}_S \triangleq \int_{0}^{B}\mathbf{S}(f)\mathbf{S}^H(f)\mathrm{d}f \geq 0$ is the auto-correlation matrix of $\mathbf{S}(f)$, and $\mathbf{R}_\Psi \triangleq  \int_{0}^{B}\mathbf{\Psi}(f)\mathbf{\Psi}^H(f)\mathrm{d}f \geq 0$ is the auto-correlation matrix of $\mathbf{\Psi}(f)$. Based on the \textit{Worst Case Uncorrelated Additive Noise} theorem from \cite{1193803}, the worst case of the distortion $\psi_i[m]$ has a zero-mean Gaussian distribution under the criterion of minimizing noise covariance. In order to make the problem analytically tractable, we assume that the distortion $\psi_i[m] \sim \mathcal{CN}(0,\sigma_{\psi}^2)$ follows circularly symmetric complex Gaussian distribution with zero mean and variance $\sigma_{\psi}^2$, which is the worst case of the distortion. The rationality of this approximation is analyzed in  Appendix.

Now, we can rewrite (\ref{e11}) as
\begin{equation}\label{e12}
	\mathbf{R}_Q = B^2\mathbf{A}\mathbf{R}_S\mathbf{A}^H + (\frac{B^2\sigma_n^2}{LT^2}+\sigma_{\psi}^2)\mathbf{I}.
\end{equation}
Since there are only $K$ channels occupied by active PUs in the spectrum $[0,W]$, $\mathbf{S}(f)$ has $K$ non-zero elements, which means that $\mathbf{R}_S$ has a rank of $K$. Perform eigendecomposition on $\mathbf{R}_Q$ as
\begin{equation}\label{e13}
	\mathbf{R}_Q = \sum_{i=1}^{p}\lambda_i\boldsymbol{v}_i\boldsymbol{v}_i^H,
\end{equation}
where $\lambda_i$ and $\boldsymbol{v}_i$ are the eigenvalues and the corresponding eigenvectors of $\mathbf{R}_Q$ respectively. Without loss of generality, we assume that the eigenvalues are arranged in descending order, i.e., $\lambda_1 \geq \lambda_2 \geq \cdots \geq \lambda_p$. Since $\mathbf{R}_S$ has a rank of $K$, and $\mathbf{A}$ is a full rank matrix, there are $p-K$ eigenvalues equal to $\frac{B^2\sigma_n^2}{LT^2}+\sigma_{\psi}^2$. As $\mathbf{R}_S$ is a positive semidefinite matrix, those $p-K$ eigenvalues must be the smallest ones. Dividing all the eigenvectors into signal subspace $\mathbf{V}_s = [\boldsymbol{v}_1,\boldsymbol{v}_2,\cdots,\boldsymbol{v}_K]$ and noise subspace $\mathbf{V}_n = [\boldsymbol{v}_{K+1}, \cdots,\boldsymbol{v}_p]$, the auto-correlation matrix $\mathbf{R}_Q$ can be denoted by
\begin{equation}\label{e14}
	\mathbf{R}_Q = \mathbf{V}_s \mathbf{\Lambda}_s \mathbf{V}_s^H + (\frac{B^2\sigma_n^2}{LT^2}+\sigma_{\psi}^2)\mathbf{V}_n\mathbf{V}_n^H,
\end{equation}
where $\mathbf{\Lambda}_s = \mathrm{diag}\{\lambda_1,\cdots,\lambda_K\}$. Define a linear system $\mathbf{U}_s = \mathbf{A}\mathbf{\Theta}_s$, $\mathbf{U}_s = \mathbf{V}_s\sqrt{\Lambda_s}$. The support of the solution of the defined linear system converges to that of (\ref{e22}). The sparsity order $K$ can be estimated by a lot of methods, such as the exponential fitting test in \cite{Angela} which is adopted in this paper. With the estimated sparsity order $\hat{K}$, the defined linear system can be easily solved by the orthogonal matching pursuit algorithm within $\hat{K}$ iterations. Because this method requires channel division, it is not suitable for systems such as spread spectrum communication systems, where all signals occupy all spectrum resources.

\section{Numerical Results}

\begin{figure}[!t]
	\centering
	\subfigure[]{\includegraphics[width=0.48\columnwidth]{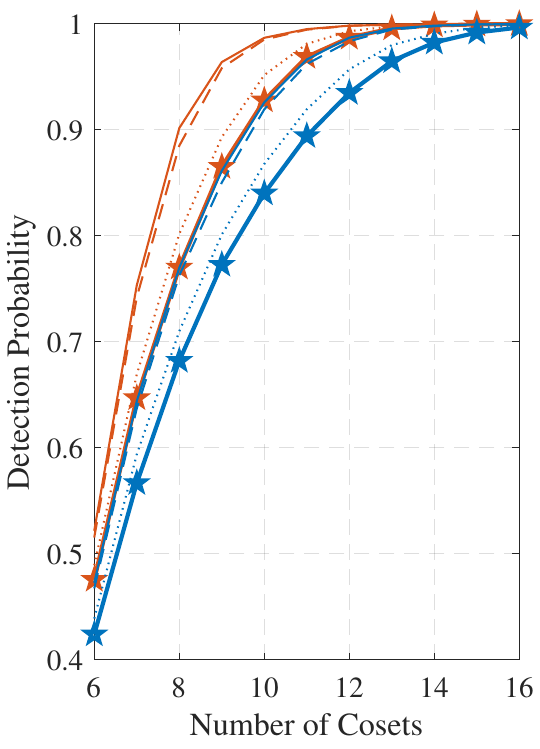}}
	\subfigure[]{\includegraphics[width=0.48\columnwidth]{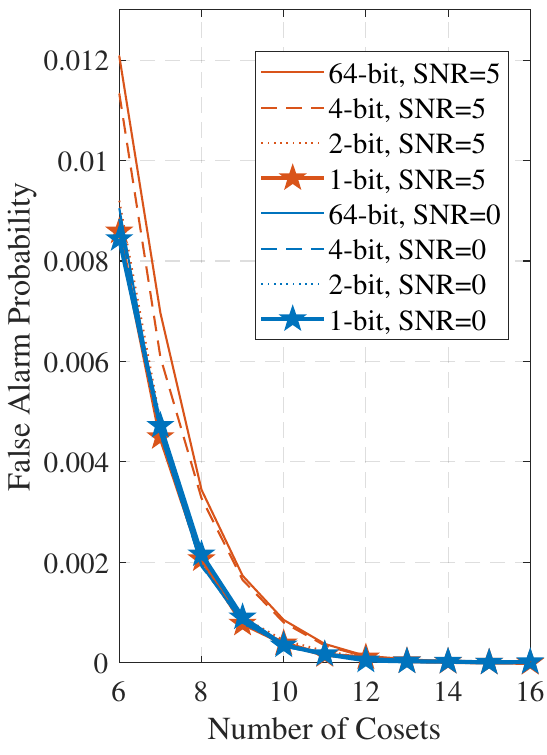}}
	\caption{Detection performance versus number of cosets under different quantization levels and SNR, with $K=4$. (a) Detection probability $P_{\text{d}}$. (b) False alarm probability $P_{\text{f}}$.}
	\label{bit}
\end{figure}

Consider a wideband signal:
\begin{equation}\label{e15}
	x(t) = \sum_{i=1}^{K} \sqrt{E_i B} s_i(t) + n(t),
\end{equation}
within the frequency range $[0, 320]$ MHz, consisting of $L=40$ channels, each with bandwidth $B = 8$ MHz. Here, $K$ is the number of channels occupied by PUs, $E_i$ represents the power coefficient of the $i$-th PU’s signal, and $n(t)$ is white Gaussian noise with zero mean and unit variance. PU signals are modeled as OFDM signals with BPSK symbols, each having $C = B/F_s = 200$ subcarriers, where $F_s$ is the symbol rate. The locations of PUs in the channels are randomly assigned. The signal-to-noise ratio (SNR) is defined as the ratio of the total power of all PU signals to the noise power. Unless specified otherwise, the default parameters in the simulation are: a time frame of $T = 25 \, \mu$s, sampling rate per coset of $f_s = B = 8$ MHz, resulting in $N = f_s T = 200$ samples per sensing frame, with time delay coefficients $\mathcal{C}$ randomly selected from $\{0,1,\dots,39\}$. 

The detection performance is evaluated by calculating the detection probability, $P_{\text{d}} = \mathbf{E} \{ {| \mathcal{S} \cap \hat{\mathcal{S}} |}/{| \mathcal{S} |} t\}$, and the false alarm probability, $P_{\text{f}} = \mathbf{E} \{ {| \hat{\mathcal{S}} \setminus \mathcal{S} |}/{L - | \mathcal{S} |} \}$, where $\left| \mathcal{S} \right|$ is the number of entries in set $\mathcal{S}$, $\hat{\mathcal{S}}$ is the estimated support set, and $\mathcal{S} \setminus \hat{\mathcal{S}}$ denotes the relative complement of $\mathcal{S}$ in $\hat{\mathcal{S}}$. Each simulation point is based on 10,000 Monte-Carlo trials.

Fig. \ref{bit} shows the detection performance of the proposed subspace-aided compressive spectrum sensing scheme with quantized samples under various quantization levels and SNRs, with $K=4$ occupied channels (10\% occupation ratio). As shown, increasing the number of cosets initially raises both the detection probability and false alarm probability. This occurs because the detection performance depends on the measurement matrix's dimensionality, which is determined by the coset count. With a low number of cosets, the samples lack sufficient information to distinguish signal from noise subspace, resulting in low detection probability and high false alarm probability. As coset count rises, detection improves, and once sufficient information is available, the false alarm probability decreases, leading to a peak in Fig. \ref{bit}(b).

The detection and false alarm probabilities for 64-bit and 4-bit quantizations are nearly identical at SNR = 5 dB and SNR = 0 dB, respectively. This suggests that 4-bit quantization is sufficient for WSS with our method, and further increases in quantization levels provide minimal performance gain. Additionally, the detection probability of 1-bit quantization at SNR = 5 dB almost matches that of 64-bit at SNR = 0 dB, with a slightly lower false alarm rate, indicating that 1-bit quantization incurs less than a 5 dB SNR loss. The quantization noise level, measured by the signal-to-quantization-noise ratio (SQNR) as $\text{SQNR} = 10\log(P_s/P_{\psi})$, where $P_s$ and $P_{\psi}$ are the powers of signal and quantization noise, respectively, rises by 6 dB per additional quantization bit \cite{6226629}. Although the theoretical SQNR loss from 1-bit to 4-bit quantization is 24 dB, Fig. \ref{bit} demonstrates that the effective SNR loss remains below 5 dB, highlighting our method's robustness in separating signal from noise components.

\begin{figure}[!t]
	\centering
	\subfigure[]{\includegraphics[width=0.48\columnwidth]{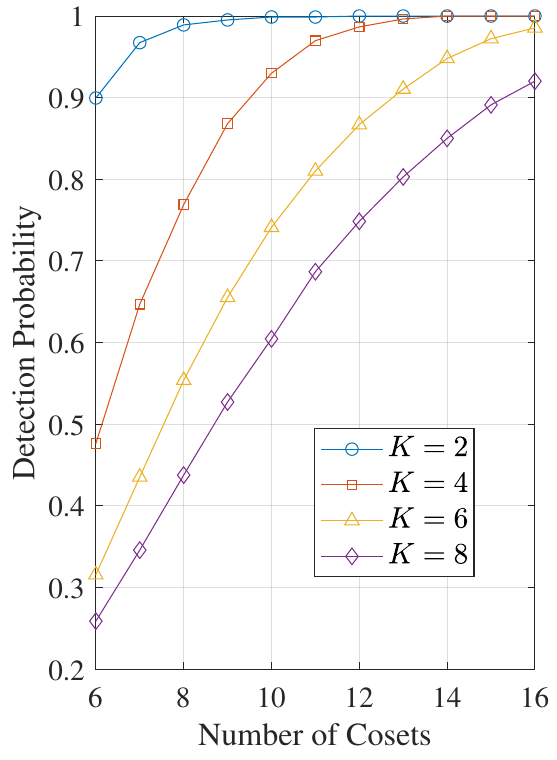}}
	\subfigure[]{\includegraphics[width=0.48\columnwidth]{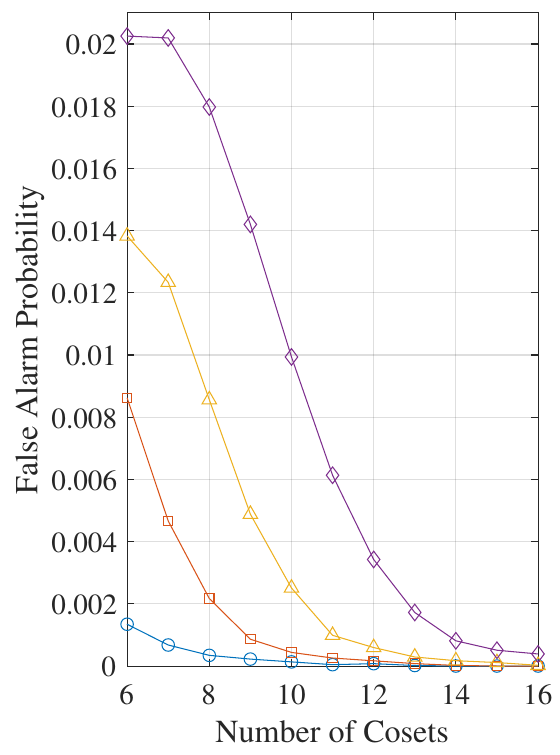}}
	\caption{Detection performance versus number of cosets under different sparsity orders $K$, with SNR = 5 dB. (a) Detection probability $P_{\text{d}}$. (b) False alarm probability $P_{\text{f}}$.}
	\label{Signalnumber}
\end{figure}

Fig. \ref{Signalnumber} illustrates the detection probability and false alarm probability of the proposed method with 1-bit quantized samples under varying sparsity orders $K$ at SNR = 5 dB. When the number of cosets reaches 20 (50\% of the Nyquist rate), detection probability approaches 100\% with near-zero false alarms across all cases, verifying the robustness of our method for different sparsity levels. Notably, for $K=2$, false alarm probability does not exhibit an initial increase due to the adequacy of 4 cosets in distinguishing two signals. As $K$ increases, more cosets are required to maintain performance, shifting the peak of the false alarm probability to higher coset counts.

\begin{figure}[!t]
	\centering
	\includegraphics[width=\columnwidth]{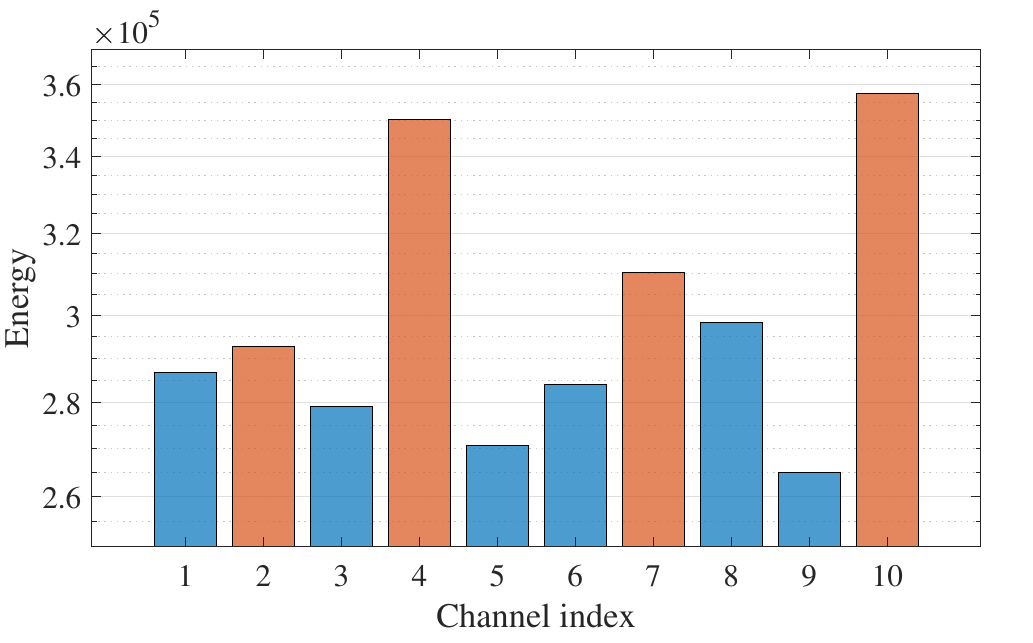}
	\caption{Power distribution of quantization noise across channels.}
	\label{power}
\end{figure}

The proposed method's primary goal is to accurately identify occupied channels despite quantization and environmental noise. Subspace decomposition helps distinguish between signal and noise components, with eigenvalues related to the channel energy. Since occupied channels have higher energy than vacant ones, eigenvalue decomposition can effectively separate the signal subspace from noise. Fig. \ref{power} shows the quantization noise power distribution across channels, revealing that quantization noise is concentrated in occupied channels and evenly spread across others, aiding in channel index estimation. In contrast, white Gaussian noise exhibits uniform energy across channels, with no such advantage for estimation. Our method, therefore, is well-suited for real quantization noise and performs even better than in the Gaussian noise case, demonstrating robust performance by approximating the worst-case noise scenario.

\section{Conclusion}

We proposed a low-power WSS method using multicoset sampling with 1-bit ADCs, significantly reducing the power demands of traditional high-precision ADCs. Our subspace-aided algorithm effectively estimates occupied channels without prior sparsity information, achieving high detection accuracy and robustness against quantization noise. Simulation results confirm that 1-bit quantization yields competitive performance with minimal SNR loss, making this approach ideal for efficient WSS in cognitive radio systems.

\bibliographystyle{IEEEtran}
\bibliography{myref}

\vfill

\end{document}